\documentclass{jpsj2} 

\title{Event-based computer simulation model of Aspect-type experiments
strictly satisfying Einstein's locality conditions\footnote{J. Phys. Soc. Jpn. 76, 104005 (2007); DOI: 10.1143/JPSJ.76.104005}}

\author{Hans De Raedt$^{1}$, Koen De Raedt$^{2}$, Kristel Michielsen$^{1}$, Koenraad Keimpema$^{1}$, and Seiji Miyashita$^{3,4}$}

\inst{
$^{1}$
Department of Applied Physics, Zernike Institute of Advanced Materials, \\
University of Groningen, Nijenborgh 4, NL-9747 AG Groningen, The Netherlands\\
$^{2}$
Department of Computer Science, University of Groningen, \\
Blauwborgje 3, NL-9747 AC Groningen, The Netherlands\\
$^{3}$Department of Physics, Graduate School of Science,\\ University of Tokyo, Bunkyo-ku,Tokyo 113-0033, Japan\\
$^{4}$ CREST, JST, 4-1-8 Honcho Kawaguchi, Saitama, Japan}

\abst{
Inspired by Einstein-Podolsky-Rosen-Bohm experiments with photons,
we construct an event-based simulation model in which every
essential element in the ideal experiment has a counterpart.
The model satisfies Einstein's criteria of local causality and
does not rely on concepts of quantum and probability theory.
We consider experiments in which the averages correspond to
those of a singlet and product state of a system of two $S=1/2$ particles.
The data is analyzed according to the experimental procedure,
employing a time window to identify pairs.
We study how the time window and the passage time of the photons,
which depends on the relative angle between their polarization
and the polarizer's direction, influences the correlations,
demonstrating that the properties of the optical elements
in the observation stations affect the correlations
although the stations are separated spatially and temporarily.
We show that the model can reproduce results which are considered to be intrinsically quantum mechanical.
} 

\recdate{November 22, 2006; accepted August 6, 2007; published October 10, 2007}

\kword{EPR paradox, computer simulation, quantum theory}

\begin{document}
\maketitle

\def\sumprime{\mathop{{\sum}'}}
\def\url#1{{\tt #1}}

\section{Introduction}
\label{intro}

Recently, there has been increasing interest in new ways of information processing
that exploit quantum mechanical correlations.
In general, quantum theory describes the state of the system by the wavefunction
from which we obtain the ensemble averaged value of quantities.
Quantum theory successfully describes
the averaged value of a quantity that is obtained in experiments by macroscopic observations.
However, when we consider quantum mechanical correlations we have to be careful.
For example, if we consider the expectation value of a product of two quantities $A$ and $B$,
we have to measure $AB$. It is not sufficient to measure $A$ and $B$ separately.
When we have to measure the correlation $AB$ experimentally,
we need a proper definition of the correlation.
In experiments, we observe $a_i$ and $b_i$ in the $i$-th measurement for $A$ and $B$.
Likewise, we must properly define the meaning of a ``pair'' of data $(a_i, b_i)$ that corresponds to $AB$.
If we overlook this point, the interesting nature of quantum correlation may disappear.
This point is most clearly illustrated by the famous Einstein-Podolsky-Rosen (EPR) paradox.

In 1935, EPR proposed a {\sl gedanken} experiment, which led them to the conclusion that quantum theory is not
a complete theory~\cite{EPR35}. Their reasoning was based on notions about completeness, physical reality
and locality.
Einstein later expressed the principle of locality as
``\textit{The real factual situation of the system $S_2$
is independent of what is done with the system $S_1$,
which is spatially separated from the former}''~\cite{BALL03}, an ontological definition of locality
that is now known as Einstein's criteria of local causality.
The question arose whether certain apparently
paradoxical predictions of quantum theory could be experimentally tested.

Bohm reformulated in 1951 the EPR {\sl gedanken} experiment into a form which is conceptually
equivalent but easier to treat mathematically~\cite{BOHM51}. In Bohm's model, a source emits pairs
of particles with opposite magnetic moments. The two particles separate spatially and propagate
in free space to an observation station in which they are detected. As the particle arrives at one
of the two observation stations it passes through a Stern-Gerlach magnet~\cite{GERL22}.
The Stern-Gerlach magnet deflects the particle, depending on the orientation of the magnet and
the magnetic moment of the particle. The deflection defines the spin ${\bf S}=\pm 1/2$ of the
particle~\cite{GERL22}. As the particle leaves the Stern-Gerlach magnet, it generates a signal
in one of the two detectors placed behind the Stern-Gerlach magnet. The firing of the detector
corresponds to a detection event.

Inspired by Bohm's proposal, Bell derived in 1964 an inequality
that imposes restrictions on the correlations
between the results of the measurements on the two spin-1/2 particles~\cite{BELL93}. Bell demonstrated that
the correlation function for the singlet state violates his inequality.
Hence, quantum theory is in conflict with at least one of the assumptions that were used in the derivation of Bell's
inequality. The demonstration of the discrepancy between certain quantum mechanical expectation values and
Bell's inequality is known as Bell's theorem~\cite{BALL03}. Bell concluded that quantum theory is not
compatible with Einstein's criteria of local causality and that no physical theory of local hidden variables
can reproduce all of the predictions of quantum theory~\cite{BELL93}.

Originally, Bell derived the inequality under the condition
that the probability distribution of the observation of $A$ is independent of that of $B$.
This is a more restrictive condition than the requirement that the physical experimental procedures
which are used to measure $A$ and $B$, are independent.
In fact, the inequality may be violated if there is some relation
between the observations of $A$ and $B$, regardless whether this relation is of quantum mechanical origin or not.
In this sense, the popular statement "a classical system cannot violate the Bell inequality" is misleading.
One has to be very careful and check if the system under study satisfies all the conditions
that are necessary to derive the inequality~\cite{LARS04,SANT05,ZUKO06}.
Indeed, it has been pointed out that under certain, physically reasonable assumptions,
a system not relying on any concept of quantum theory and that obeys Einstein's criteria of local causality
can also violate the original Bell inequality~\cite{LARS04,RAED06c}.
The common feature of these models is the presence of a time window to identify the single two-particle systems,
as in real EPRB experiments~\cite{FREE72,ASPE82b,ASPE82a,TAPS94,TITT98,WEIH98,FATA04,SAKA06}.
From these observations, it is clear that a violation of Bell's inequality
is not enough to conclude that there are quantum correlations~\cite{LARS04,RAED06c}.
In this paper, we study the two-particle correlations with one of these, what might be called, classical models
and we demonstrate that their key feature,
the dependence of the correlation on the time window that is used to identify the pair,
allows us to reproduce the correlations that are characteristic for a quantum system
of two $S=1/2$ particles.


\begin{figure*}[t]
\begin{center}
\includegraphics[height=10cm]{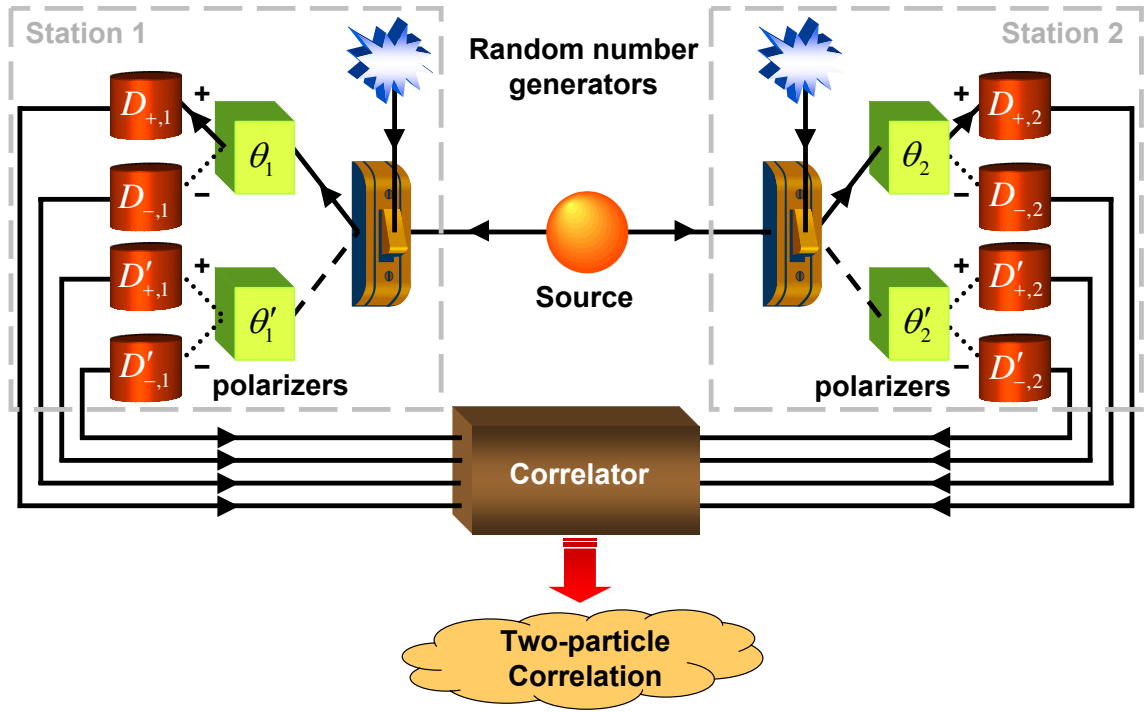}
\caption{(color online)
Schematic diagram of an EPRB experiment with photons~\cite{ASPE82b}.
}
\label{aspect}
\label{fig1}
\end{center}
\end{figure*}


For this purpose, we construct a computer model
that satisfies Einstein's criteria of local causality
and use this model to simulate the experiment with optical switches,
as performed by Aspect \textit{et al.}~\cite{ASPE82b}.
The sources used in EPRB experiments with photons emit photons with opposite but otherwise
unpredictable polarization. We refer to this experimental set-up as Experiment I.
Inserting polarizers between the source and the
observation stations changes the pair generation procedure such that
the two photons have a fixed polarization. We refer to this set-up
as Experiment II.
As a result of the fixed polarization of the photons the photon intensity measured in the detectors
behind the polarizers in each observation station obeys Malus' law.
Our simulation model reproduces the correct quantum mechanical behavior for the
single-particle and two-particle correlation function for both types of experiments.
The difference between this model and the model described in Ref.~\cite{RAED06c}, is the algorithm
to simulate the polarizer. In Ref.~\cite{RAED06c} we used a model for the polarizers that is too simple to
correctly describe experiments of type II.

This paper is organized as follows.
In Section~\ref{sec2} we describe the experimental set-up, the data gathering method and the data
analysis procedures used in EPRB experiments with photons, and in particular in the timing experiment
with optical switches by Aspect \textit{et al.}~\cite{ASPE82b}.
In Appendix A we present an analysis of real experimental data of another EPRB experiment with photons~\cite{WEIH98}
that is the successor of the experiment by Aspect \textit{et al.}~\cite{ASPE82b}.
We make a distinction between
experiments of type I and II. A brief review of the analysis of the experiments in the framework of quantum theory is
given in Section~\ref{sec3}.
We give explicit expressions for the single- and two-particle expectation values
for both types of experiments and we introduce Bell's inequality. We discuss the fundamental problem of
relating quantum theory with the data set recorded in the experiment.
In Section~\ref{sec4} we describe our computer simulation model of EPRB experiments with photons.
We give an explicit description of the algorithm to simulate the photons, the observation stations containing
the polarizers and detectors, and the data analysis procedure. Every essential element of the experiment has a counterpart
in the algorithm. For some model parameters we can compute the two-particle correlation function analytically,
as shown in Appendix B.
In Section~\ref{sec5} we discuss our simulation and analytical results and where appropriate we
compare them to the results obtained from quantum theory.
Section~\ref{sec6} presents a summary and a discussion of our results.

\section{EPRB experiment with photons}\label{sec2}

A schematic diagram of the type I timing experiment with optical switches
is shown in Fig.~\ref{fig1} (see also Fig.~2 in~\cite{ASPE82b}).
A source emits pairs of photons with
opposite polarization.
Each photon of a pair propagates to an observation station
in which it is manipulated and detected.
The two stations are separated spatially and temporally.
This arrangement prevents the observation at
station 1 (2) to have a causal effect on the
data registered at station $2$ (1).

As the photon arrives at station $i=1$ and $i=2$, it passes through an optical switch
that directs the photon to one of the two polarizers with a fixed orientation (see Fig.~\ref{fig1}).
The orientation of the two polarizers in each observation station is characterized by the angles $\theta_i$ and
$\theta_i^{\prime}$.
As the photon leaves the polarizer, it generates a signal in one of the
two detectors.
Each station has it own clock that assigns a time-tag
to each signal generated by one of the two detectors.
Effectively, this procedure discretizes time in intervals of a width that is
determined by the time-tag resolution $\tau$.
The time-tag generators are synchronized before each run. This
procedure is necessary because in time, the clocks may become unsynchronized.
Furthermore, in real experiments, some photons may not be detected.
In this paper, we consider ideal experiments only.
Hence, our simulation procedure does not allow for practical loopholes, such as the detection loophole
(lost photons and registration of accidental signals), the ``fair sampling'' loophole,
and synchronization problems, that might be present in real optical experiments.

In the experiment, the firing of a detector
is regarded as an event.
At the $n$th event, the data recorded on a hard disk (not shown) at station $i=1,2$
consists of $\gamma_{n,i}=\theta_{n,i},\theta_{n,i}^{\prime}$, depending on the state of the
optical switches,
$x_{n,i}=\pm 1$, specifying
which of the two detectors behind the selected polarizer fired and
the time tag $t_{n,i}$
indicating the time at which a detector fired.
Hence, the set of data collected at station $i=1,2$ during a run of $N$ events
may be written as
\begin{align}
\label{Ups}
\Upsilon_i=\left\{ {x_{n,i} =\pm 1,t_{n,i},\gamma_{n,i} \vert n =1,\ldots ,N } \right\}
.
\end{align}

Any experimental procedure requires some criterion to decide which detection
events are to be considered as stemming from a single two-particle system.
In EPRB-type experiments with photons, this decision is taken on the basis of coincidence in time~\cite{WEIH98,CLAU74}.
Coincidences are identified by comparing the time differences
$\{ t_{n,1}-t_{n,2} \vert n =1,\ldots ,N \}$ with a time window $W$~\cite{WEIH98}. 
Thus, for each pair of rotation angles $\alpha=\theta_1,\theta_1^{\prime}$ and $\beta=\theta_2,\theta_2^{\prime}$,
the number of coincidences between detectors $D_{x,1}$ ($x =\pm $1), $D_{x,1}^{\prime}$ ($x =\pm $1) at station
1 and detectors $D_{y,2}$ ($y =\pm $1), $D_{y,2}^{\prime}$ ($y =\pm $1) at station 2 is given by
\begin{align}
\label{Cxy}
C_{xy}=C_{xy}(\alpha,\beta)=&
\sum_{n=1}^{N}
\delta_{x,x_{n ,1}} \delta_{y,x_{n ,2}} \delta_{\alpha ,\gamma_{n,1}}
\delta_{\beta,\gamma_{n,2}}
\Theta(W-\vert t_{n,1} -t_{n ,2}\vert)
,
\end{align}
where $\Theta (t)$ is the Heaviside step function.
The correlation $E(\alpha,\beta)$ between the coincidence counts
is then given by
\begin{align}
\label{Exy}
E(\alpha ,\beta )&=
\frac{C_{++}+C_{--}-C_{+-}-C_{-+}}{C_{++}+C_{--}+C_{+-}+C_{-+}}
,
\end{align}
where the denominator in Eq.(\ref{Exy}) is the sum of all coincidences.
In practice, the data $\{\Upsilon_1,\Upsilon_2\}$ are analyzed
long after the data has been collected.
In general, the numerical values for the coincidences
$C_{xy}(\alpha,\beta)$ and correlation $E(\alpha ,\beta )$ depend on the time-tag resolution
and the time window used to identify the coincidences.

In Experiment II, extra polarizers are inserted between the source
and the observation stations~\cite{ASPE82b}.
We denote the orientations of these polarizers by the angles $\eta_1$ and $\eta_2$.

\subsection{Role of the time window}\label{sec3a}

As we already mentioned, in our simulation we leave no room for practical loopholes such as detection
and "fair sampling" loopholes that may be used to invalidate the conclusions drawn from real experiments.
The point of view taken in this paper is that we want to perform a simulation of ideal
experiments and show that we can reproduce the results of quantum theory.

Most theoretical treatments of the EPRB experiment assume that the correlation,
as measured in experiment, is given by~\cite{BELL93}
\begin{align}
\label{CxyBell}
C_{xy}^{(\infty)}&=\sum_{n=1}^N\delta_{x,x_{n ,1}} \delta_{y,x_{n ,2}}
,
\end{align}
%
%
%
where we assume that the pairs are well defined.
This expression, however, is obtained from  Eq.~(\ref{Cxy})
by taking the limit $W\rightarrow\infty$, hence the notation $C_{xy}^{(\infty)}$.
An argument that might justify taking the limit $W\rightarrow\infty$ and hence the
expectation that the correlation does not strongly depend on $W$
(disregarding statistical fluctuations), is the hypothesis that the time differences originate
from some random processes that do not depend on the polarization of the photons and on the
settings of the polarizers. However, the assumption that the time differences are independent random
variables may not be correct and in fact, in experiments,
a lot of effort is made to reduce (not increase) $W$~\cite{WEIH98} (see also Appendix A).
As we will see later, our simulation results
agree with the results of quantum theory if
we assume that the time differences are random variables
that depend on the settings of the polarizers and the polarization
and if we consider the limit $W\rightarrow0$.

\section{Quantum Theory}\label{sec3}
\label{sec:quantum}

As is well known, quantum theory itself has nothing to say about the individual events
as they are observed in experiments (quantum measurement paradox),
but it provides a framework to compute the probability
for the various possible events to occur~\cite{BALL03,HOME97}.
In this section, we give a brief account of the quantum mechanical calculation
of the averages obtained in the EPRB experiment described earlier, strictly staying
within the axiomatic framework that quantum theory provides.

In the quantum mechanical description of Experiment I,
the source is assumed to emit two photons
of which the polarization is described by the state
\begin{align}
\label{eq7}
| \Psi \rangle &=\frac{1}{\sqrt 2 }\left( {| H \rangle
_1 | V \rangle _2 -| V \rangle _1 | H
\rangle _2 } \right)
=\frac{1}{\sqrt 2 }\left( {| {HV}
\rangle -| {VH} \rangle } \right),
\end{align}
where $H$ and $V$ denote the horizontal
and vertical polarization and the subscripts refer to photon 1 and 2, respectively.
The state $|\Psi\rangle$ cannot be written as a
product of single-photon states, hence it is an entangled state.

In Experiment II, the photons have a definite polarization when
they enter the observation station.
The polarization of the two photons is described by
the product state
\begin{align}
\label{eq23}
|\Psi\rangle =&(\cos \eta_1|H\rangle_1 +\sin \eta_1|V\rangle_1)
(\cos \eta_2 |H\rangle_2 +\sin \eta_2 |V\rangle _2).
\end{align}

Each of the polarizers in the observation stations splits the beam of incoming photons.
Using the fact that the two-dimensional vector space with basis vectors $\{|H\rangle,|V\rangle\}$
is isomorphic to the vector space of spin-1/2 particles,
we may use the quantum theory of the latter to describe the action of a polarizer
as a rotation about its angle of orientation ($\theta_1$, $\theta_1^{\prime},\theta_2$ or $\theta_2^{\prime}$),
followed by the measurement of the $z$-component of the Pauli spin matrix.
More specifically, a polarizer with orientation $\alpha$
changes the states $| H \rangle$ and $| V\rangle$
according to
\begin{align}
\label{eq8}
 | H \rangle &\to  \phantom{-}\cos \alpha | H \rangle +\sin \alpha | V \rangle,
\nonumber \\
 | V \rangle &\to  -\sin \alpha | H \rangle +\cos \alpha | V \rangle.
 \end{align}
Hence, the polarizers at station 1 and 2 with orientation $\alpha=
\theta_1, \theta_1^{\prime}$ and
$\beta=\theta_2,\theta_2^{\prime}$, respectively, change the state $|\Psi\rangle$ into
\begin{equation}
\label{eq9}
| \Phi \rangle =R(\alpha )R(\beta )| \Psi \rangle ,
\end{equation}
where it is implicitly understood that $R(\alpha )$  and
$R(\beta )$ operate on the spin of particle 1 and 2, respectively.
The rotation matrix $R(\theta )$ is given by
\begin{equation}
\label{eq10}
R(\theta )=\left(
\begin{array}{rr}
 \phantom{-}\cos\theta & \phantom{-}\sin\theta\\
 -\sin\theta & \phantom{-}\cos\theta
\end{array}
\right).
\end{equation}
According to quantum theory, the expectation value of counting
photons at the $+$ ($-$) detector behind the polarizer with orientation $\alpha=\theta_1, \theta_1^{\prime}$
($\beta=\theta_2, \theta_2^{\prime}$)
is given by~\cite{BALL03}
\begin{align}
\label{eq11}
P_+(\alpha)&= \langle \Phi |1+\sigma _1^z | \Phi \rangle/2
=1/2+\langle \Psi |R^{-1}(\alpha )\sigma _1^z R(\alpha )| \Psi
\rangle/2
\nonumber \\&
=1/2+\langle \Psi |\sigma _1^z \cos 2\alpha +\sigma
_1^x \sin 2\alpha | \Psi \rangle/2,
\nonumber \\
P_-(\beta)&= \langle \Phi |1-\sigma _2^z| \Phi \rangle/2
=1/2-\langle \Psi |R^{-1}(\beta )\sigma _2^z R(\beta )| \Psi
\rangle/2
\nonumber \\
&=1/2-\langle \Psi |\sigma _2^z \cos 2\beta +\sigma _2^x
\sin 2\beta | \Psi \rangle/2,
\end{align}
where $\sigma _1 =(\sigma _1^x ,\sigma _1^y ,\sigma _1^z )$ and $\sigma _2
=(\sigma _2^x ,\sigma _2^y ,\sigma _2^z )$ are the Pauli spin-1/2 matrices
for particles 1 and 2, respectively~\cite{BALL03}.

The expectation values of the $z$-components of the Pauli-spin
matrices are given by
\begin{align}
\label{eq12a}
E_1(\alpha)&=  \langle \Phi |\sigma _1^z | \Phi \rangle=P_+(\alpha)-P_-(\alpha) ,
\nonumber \\
E_2(\beta)&=  \langle \Phi |\sigma _2^z | \Phi \rangle=P_+(\beta)-P_-(\beta) .
\end{align}
The main objective of EPRB experiments is to measure the two-particle
correlation
\begin{align}
\label{eq12}
 E(\alpha ,\beta )&=  \langle \Phi |\sigma _1^z \sigma _2^z
| \Phi \rangle
=
\langle\Psi |R^{-1}(\alpha )\sigma _1^z R(\alpha )R^{-1}(\beta )\sigma _2^z R(\beta )| \Psi
\rangle.
\end{align}
Table I gives the explicit expressions for
the expectation values defined by Eqs.~(\ref{eq11}),~(\ref{eq12a}) and~(\ref{eq12})
for the two different types of experiments.
\begin{table}
\caption{The single- and two-particle expectation values
defined by Eqs.~(\ref{eq11}),~(\ref{eq12a}) and~(\ref{eq12}) for the
two experiments described by the states Eqs.~(\ref{eq7}) and~(\ref{eq23}), respectively. }
\label{tab:1}       
\begin{tabular}{rrr}
\hline\noalign{\smallskip}
& Experiment I & Experiment II  \\
\noalign{\smallskip}\hline\noalign{\smallskip}
$P_+(\alpha)$ & $1/2$ & $\cos^2(\alpha-\eta_1)$\\
$P_-(\beta)$ & $1/2$ & $\sin^2(\beta-\eta_2)$\\
$E_1(\alpha)$ & $0$ & $\cos2(\alpha-\eta_1)$\\
$E_2(\beta)$ & $0$ & $\cos2(\beta-\eta_2)$\\
$E(\alpha,\beta)$ & $-\cos2(\alpha-\beta)$ & $\cos 2(\alpha-\eta_1) \cos 2(\beta-\eta_2)$\\
\noalign{\smallskip}\hline
\end{tabular}
\end{table}
From Table I, it is clear that
measuring $E_1(\alpha)$, $E_2(\beta)$ and $E(\alpha,\beta)$
for various $\alpha$ and $\beta$
suffices to distinguish between systems in
the entangled state (Experiment I) or in the product state (Experiment II).

Data of EPRB experiments are often analyzed in terms of the function~\cite{WEIH98,CLAU69}
\begin{align}
\label{eq29}
S(\alpha ,{\alpha }',\beta ,{\beta }')=&{E(\alpha ,\beta )-E(\alpha
,{\beta }')}
+{E({\alpha }',\beta )+E({\alpha }',{\beta }')},
\end{align}
because it provides clear evidence that
the system is described by an entangled state.
The idea behind this reasoning is that for any product state
\begin{equation}
\label{eq30}
-2\le S(\alpha, \alpha',\beta,\beta')\le 2,
\end{equation}
an inequality known as one of Bell's generalized inequalities~\cite{CLAU69}.
This can be seen as follows.
For any product state $|\Psi\rangle$, we have $E(\alpha,\beta)=E_1(\alpha)E_2(\beta)$.
Let us denote $a=E_1(\alpha)$, $b=E_1(\alpha')$,
$c=E_2(\beta)$, and $d=E_2(\beta')$.
Clearly, $a,b,c,d\in[-1,1]$. For any $a,b,c,d\in[-1,1]$ we have~\cite{ACCA05}
\begin{align}
\label{eq30a}
|ac-ad+bc+bd|&\le |ac-ad|+|bc+bd|
\le |a||c-d|+|b||c+d|
\nonumber \\&
\le |c-d|+|c+d|
\le 1-cd+1+cd
\nonumber \\
&\le 2,
\end{align}
hence Eq.~(\ref{eq30}) follows.
Thus, we conclude that if $|\Psi\rangle$ can be written as a product state,
we must have
\begin{align}
S_{max}\equiv\max_{\alpha,\alpha',\beta,\beta'}|S(\alpha,\alpha',\beta,\beta')|\le 2.
\end{align}
Furthermore, it can be shown that~\cite{CIRE80}
\begin{align}
\label{eq30b}
|S(\alpha, \alpha',\beta,\beta')|\le2\sqrt{2},
\end{align}
independent of the choice of $|\Psi\rangle$.
In other words, if
%
$2<S_{max}\le 2\sqrt{2}$,
the quantum system is in an entangled state.

For later use, we introduce the function
\begin{equation}
\label{eq31}
S(\theta )\equiv S(\alpha ,\alpha +2\theta ,\alpha +\theta ,\alpha +3\theta),
\end{equation}
where we have fixed the relation between the angles $\beta =\alpha
+\theta $, ${\alpha }'=\alpha +2\theta $, ${\beta }'=\alpha +3\theta $
through the angle $\theta $.
Because of rotational invariance, $S(\theta )$ does not depend on $\alpha $
and therefore, we set $\alpha=0$ to simplify matters a little.
In the case of Experiment I, $E(\alpha ,\beta )=-\cos 2(\alpha -\beta )$
and we find
\begin{equation}
\label{Stheta}
S(\theta )=3\cos 2\theta -\cos 6\theta,
\end{equation}
which reaches its maximum value $S_{max}=\max _\theta S(\theta )=2\sqrt 2$
at $\theta =\pi /8+j\pi /2$, where $j$ is an integer number.

Analysis of the experimental data~\cite{FREE72,ASPE82b,ASPE82a,TAPS94,TITT98,WEIH98,ROWE01,FATA04,SAKA06},
yields results that are in good agreement with the expressions in Table I,
leading to the conclusion that in a quantum mechanical description of Experiment I,
the state does not factorize, in spite of the fact that the particles are
spatially and temporally separated and do not interact.
Our analysis of the few experimental data for Experiment I
that is publicly available~\cite{WEIHdownload} supports this conclusion (see Appendix A).

\subsection{From quantum theory to data}
According to the formalism of quantum theory, the result of a measurement
is an eigenvalue of the dynamical variable that is being measured~\cite{BALL03}.
Applied to the case of the EPRB experiment, each measurement yields
an eigenvalue of the matrices
$A=R^{-1}(\alpha )\sigma _1^z R(\alpha )$,
$B=R^{-1}(\beta )\sigma _2^z R(\beta )$,
and
$C=R^{-1}(\alpha )R^{-1}(\beta )\sigma _1^z \sigma _2^z R(\beta )R(\alpha )$.
Obviously, the $n$th measurement of $A$, $B$, or $C$
yields an eigenvalue $a_n=\pm1$, $b_n=\pm1$, or $c_n=\pm1$, respectively.

The conventional interpretation of quantum theory asserts that the outcome of
each measurement constitutes a Bernoulli trial,
that is we assign the same probability to an outcome,
independent of which trial is considered and independent of what
happened in any of the other measurements.
In other words, the probability to observe for instance $a_n$ is logically
independent from the probability to observe $a_m$ for all $n\not=m$.

For simplicity, we now focus on the case where all photons are directed towards
the polarizers with orientation $\theta_i$.
Let us then inquire how we can simulate the quantum mechanical results
of the EPRB experiment (see Table I) without leaving the framework of quantum theory.
Evidently, this is a nearly trivial exercise.
All we have to do is set up three Bernoulli processes that
generate sets of data $\{a_n=\pm1,b_n=\pm1,c_n=\pm1|n=1,\ldots,N\}$
such that
\begin{align}
\frac{1}{N}\sum_{n=1}^N a_n
&\approx
E_1(\theta_1)
,\quad
\frac{1}{N}\sum_{n=1}^N b_n
\approx
E_2(\theta_2)
,\quad
\frac{1}{N}\sum_{n=1}^N c_n
\approx
E(\theta_1,\theta_2),
\end{align}
for all $\theta_i$ and large $N$.
However, this line of reasoning brings out the fundamental problem of relating
the set of data
\begin{align}
{\cal Q}=\{a_n,b_n,c_n|n=1,\ldots,N\},
\end{align}
obtained from quantum theory with the set of data
\begin{align}
{\cal E}&=\{ x_{n,1},x_{n,2},t_{n,1},t_{n,2} \vert n =1,\ldots ,N \},
\end{align}
recorded in the experiment.

In the case that we measure a property of a single particle,
we may identify $x_{n,1}$ with $a_{n}$ and $x_{n,2}$ with $b_{n}$, yielding
\begin{align}
\frac{1}{N}\sumprime_{n=1}^N x_{n,1}
&\approx
E_1(\theta_1)
,\quad
\frac{1}{N}\sumprime_{n=1}^N x_{n,2}
\approx
E_2(\theta_2),
\end{align}
where the prime indicates that the sum runs over all events that yield a coincidence,
that is for all events for which $\Theta(W-\vert t_{n,1} -t_{n ,2}\vert)=1$.

As we know from the work of Bell and others~\cite{BELL93,SICA99},
simple assignments of the form $c_n\leftrightarrow x_{n,1}x_{n,2}$ cannot reproduce the result of
quantum theory for $E(\theta_1,\theta_2)$.
However, as we need to identify pairs for measurements that involve properties of two particles,
there is no simple a-priori rule to relate
$c_n$ to the data $\{ x_{n,1},x_{n,2},t_{n,1},t_{n,2}\}$.
In EPRB experiments with photons the correlation $E(\theta_1,\theta_2)$ is calculated according to Eq.~(\ref{Exy}),
using the coincidences Eq.~(\ref{Cxy}).
That is, we adopt the assignment
\begin{align}
\label{cn}
c_n\leftrightarrow\frac{x_{n ,1}x_{n ,2} \Theta(W-\vert t_{n,1} -t_{n ,2}\vert)}{
\sum_{n=1}^{N}\Theta(W-\vert t_{n,1} -t_{n ,2}\vert)}N.
\end{align}
In Eq.~(\ref{cn}), the coincidence window $W$ enters because it is necessary to have a criterion
to decide which particles belong to a single two-particle system, an essential ingredient in any real
EPRB experiment. As this choice is, in a sense, ad hoc, we have to study the significancy and
implications of this choice, as we will do in the next sections.
%

\section{Simulation model}\label{sec4}
\label{sec:computer}

We now take up the main challenge, the construction of processes
that generate the data sets Eq.~(\ref{Ups}) such that
they reproduce the results of quantum theory, summarized in Table I.
A concrete simulation model of the EPRB experiment sketched in Fig.~\ref{fig1} requires
a specification of the information carried by the particles,
of the algorithm that simulates the source and
the observation stations, and of the procedure to analyze the data.
From the specification of the algorithm, it will be clear that it complies with
Einstein's criteria of local causality on the ontological level: Once the particles
leave the source, an action at observation station 1 (2) can, in no way,
have a causal effect on the outcome of the measurement at observation
station 2 (1).

\subsection{Source and particles}
The source emits particles that carry a vector
${\bf S}_{n,i}=(\cos(\xi_{n}+(i-1)\pi/2) ,\sin(\xi_{n}+(i-1)\pi/2))$,
representing the polarization of the photons.
The ``polarization state'' of a particle is completely characterized by $\xi _{n}$,
which is distributed uniformly over the interval $[0,2\pi[$.
We use uniform random numbers
to mimic the apparent unpredictability of the experimental data.
However, from the description of the algorithm, it trivially follows that
instead of uniform random number generators, simple counters that sample
the interval $[0,2\pi[$ in a systematic, but uniform, manner might be employed
as well. This is akin to performing integrals by the trapezium rule instead
of by Monte Carlo sampling.

\subsection{Observation stations}

The input-output relation of a polarizer is rather simple: For each input
event, the algorithm maps the input vector \textbf{S} onto a single output
bit $x$. The value of the output bit depends on the orientation of the
polarizer ${\bf a}=(\cos\alpha,\sin\alpha)$. According to Malus' law, for fixed
${\bf S}=(\cos\xi,\sin\xi)$ and fixed ${\bf a}$, the bits $x_{n}$ are to be generated such
that
\begin{equation}
\label{eq25}
\mathop {\lim }\limits_{N\to \infty }
\frac{1}{N}\sum\limits_{n=1}^N {x_n }=\cos2(\xi-\alpha),
\end{equation}
with probability one.
If, as in Experiment I, the input vectors \textbf{S} are distributed
uniformly over the unit circle, the sequence of output bits should satisfy
\begin{equation}
\label{eq26}
\mathop {\lim }\limits_{N\to \infty } \frac{1}{N}\sum\limits_{n=1}^N {x_n }
=0,
\end{equation}
with probability one,
independent of the orientation \textbf{a} of the polarizer.

As we work under the hypothesis of ideal experiments,
the algorithm to simulate each of the four different polarizers in Fig.~\ref{aspect}
should be identical.
Evidently, for the present purpose,
if we switch from Experiment I to Experiment II, it is not
permitted to change the algorithm for the polarizer.

In this paper, we use a deterministic model for a polarizer.
Elsewhere, we have demonstrated that simple deterministic, local, causal and classical
processes that have a primitive form of learning capability
can be used to simulate quantum systems, not by solving a wave equation
but directly through event-by-event simulation~\cite{RAED05b,RAED05c,RAED05d,MICH05,RAED06a}.
The events are generated such that their frequencies of occurrence agree with the quantum mechanical probabilities.
In this simulation approach, the basic processing unit is called a
deterministic learning machine (DLM)~\cite{RAED05b,RAED05c,RAED05d,MICH05,RAED06a}.
A DLM learns by processing successive events but does not store the data contained in the individual events.
Connecting the input of a DLM to the output of another DLM yields
a locally connected network of DLMs.
A DLM within the network locally processes the data contained
in an event and responds by sending a message that may be used as input for another DLM.
DLMs process messages in a sequential manner and
only communicate with each other by message passing.
In a simple physical picture, a DLM is a device that
exchanges information with the particles that pass through it.
It learns by comparing the message carried by an event with predictions
based on the knowledge acquired by the DLM during the processing of previous events.
The DLM tries to do this in an efficient manner, effectively by minimizing the difference
of the data in the message and the DLM's internal representation of it~\cite{RAED05b,RAED05c,RAED05d,MICH05,RAED06a}.
We now describe the DLM that simulates the operation of a polarizer~\cite{RAED05b,RAED06a}.

Let us focus on the polarizer with orientation $\theta_1$.
The DLM has an internal two-dimensional unit vector ${\bf R}_n =(x_n,y_n)$
and a parameter $0<l<1$ that controls the pace of learning (to be discussed later).
The orientation ${\bf a}=(\cos \theta_1 ,\sin \theta_1 )$
of the polarizer is also part of the input to the DLM.
The DLM receives as input, the sequence of unit vectors
${\bf S}_n =(\cos \xi _n ,\sin \xi _n )$ for $n=1,\ldots ,N$
where $N$ is the total number of events, generated by the source.
We know that the output signal of a polarizer depends on the difference
between the polarization and the orientation of the polarizer only (Malus' law).
This is taken into account by rotating ${\bf S}_{n}$ about $\theta_1 $.
We denote the resulting vector by
${\bf Y}_n =(\cos (\xi _n -\theta_1 ),\sin (\xi _n -\theta_1 ))$.
For each input event, the DLM computes eight trial vectors
according to the following rules~\cite{RAED05b,RAED06a}
\begin{align}
\label{eq27}
 \hat x_n &= l{s}'x_n\frac{1+\Delta}{2}+s\sqrt {1-l^2+l^2x_n^2 }\frac{1-\Delta}{2},
 \nonumber \\
 \hat y_n &= l{s}'y_n\frac{1-\Delta}{2}+s\sqrt {1-l^2+l^2y_n^2 }\frac{1+\Delta}{2},
\end{align}
where $s,{s}'=\pm 1$ are variables that allow us to generate the trial vectors
in each of the four quadrants and $\Delta=\pm1$ determines whether
the trial vector is obtained by rescaling the $x$-coordinate ($\Delta=1$)
or the $y$-coordinate ($\Delta=-1$). Note that each of the eight rules
generates a unit trial vector.

The final step in the DLM algorithm consists of comparing
the vector ${\bf Y_n}$ with each of the eight trial vectors  $(\hat x_n,\hat y_n)$.
The DLM updates its internal vector ${\bf R}_n$ by choosing the trial vector,
that is the triple $(\Delta ,s,{s}')$, for
which the Euclidian distance $\Vert{\bf Y}_{n}-(\hat x_n,\hat y_n)\Vert$ is minimal.
If the trial vector was obtained by applying a $\Delta =+1$ rule,
the DLM generates a $+1$ output event. Otherwise it generates a $-1$ output event.

Elsewhere, we have shown by means of simulations and analytical
methods that the DLM described above generates $-1$ and $+1$ events,
that are distributed according to Malus' law if the input vector
${\bf S}_n =(\cos \xi_n,\sin \xi_n )$ does not change during a sufficiently
long sequence of events~\cite{RAED05b,RAED06a}.
If the input vector ${\bf S}_n =(\cos \xi _n ,\sin \xi _n )$ is uniformly
distributed over the unit circle, the dynamics of the DLM
generates events according to the function $\hbox{sign}(\cos 2(\xi_n-\theta_1))$
(results not shown).

The DLM is a machine with elementary learning capabilities: It is
an adaptive system that learns from the input events~\cite{RAED05b,RAED05c,RAED06a}.
The parameter $0<l<1$ controls the speed of the
learning process and the accuracy with which the internal vector can
represent input vectors. If $l$ is close to one, the DLM learns slow and gives
an accurate response but it also ``forgets'' slow, that is if the DLM is
offered different input vectors, it may take long before its internal vector
has adapted to the new situation. In a sense, $l$ controls the ``coherence'' of
the system~\cite{RAED05b,RAED05c,RAED06a}.
As the DLM, operating according to the rules Eq. (\ref{eq27}),
can reproduce both Malus' law and the function  $\hbox{sign}(\cos 2(\xi_n-\theta_1))$,
all that is left to do to completely specify the model of the polarizer is
to add the mechanics for the time tagging.

To assign a time-tag to each event,
we assume that as a particle passes through the detection system, it may experience a time delay.
This is a key assumption in the construction of the simulation model.
In principle, the time-delay of the individual photons cannot be
derived from the Maxwell equations because they describe waves, not particles.
Thus, to find an a-priori justification for the assumption that the particle experiences
a time-delay we are limited to making inferences from experimental data.
Empirical evidence is provided in Appendix A where we analyze experimental data
of an EPRB experiment with photons and demonstrate that the average time-of-flight of the photons
depends on the orientation of the polarizer.
Thus, there is experimental evidence that supports the assumption
that the photons experience a time delay as they pass through the polarizers.
It is unfortunate that the experimental data that is publicly available is far too
scarce to allow a detailed analysis of the time-delay mechanism.
Therefore, as a model of the time delay, we will choose a specific model that is as simple as possible,
is in concert with empirical knowledge, is capable of reproducing
the results of quantum theory and allows an analytical treatment in particular limiting cases.

In our model of the time delay, ${t}_{n ,i} $ for a particle
is assumed to be distributed uniformly over the interval $[t_{0}, t_{0}+T]$.
In practice, we use uniform random numbers
to generate ${t}_{n ,i}$.
As in the case of the angles $\xi_{n}$, the random choice of ${t}_{n ,i}$
is merely convenient, not essential.
From Eq.(\ref{Cxy}), it follows that only differences of time delays matter.
Hence, we may put $t_0=0$.
The time-tag for the event $n$ is then $t_{n,i}\in[0,T]$.

We now come to the point that we have to specify $T$ explicitly.
In fact, there are not many options.
Is is an experimental fact that the output intensities
of the two light beams emerging from a polarizer depend
on the direction of polarization of the incident light,
relative to the angle of the main optical axis of the polarizer crystal~\cite{BORN64}.
Thus, at least macroscopically, the system consisting of a polarized light beam and polarizer
is invariant for rotations about the direction of propagation of the light wave.
Assuming that this invariance carries over to the individual particles,
we can construct only one number that depends on the relative angle: ${\bf S}_{n} \cdot {\bf a}$,
implying that $T=T(\xi _{n } -\theta_1 )$ can depend on $\xi _{n } -\theta_1 $ only.
Furthermore, consistency with classical electrodynamics requires that
functions that depend on the polarization have period $\pi$~\cite{BORN64}.
Thus, we must have $T(\xi _{n } -\theta_1)=F( ({\bf S}_{n ,1} \cdot {\bf a})^2)$.
Of course, the arguments that have been used to arrive at this form are not sufficient
to uniquely fix the form of $F( ({\bf S}_{n ,1} \cdot {\bf a})^2)$.
As explained earlier, the available experimental data does not suffice
to determine the form of $F( ({\bf S}_{n ,1} \cdot {\bf a})^2)$.
Therefore we use simplicity as a criterion to select a specific form.
By trial and error, we found that $T(\xi _{n } -\theta_1 )=T_0 F(|\sin 2(\xi _{n } -\theta_1)|)=T_0|\sin 2(\xi _{n } -\theta_1)|^d$
yields useful results.
Here, $T_0 =\max_\theta T(\theta)$ is the maximum time delay
and defines the unit of time, used in the simulation and $d$ is a free parameter of the model.
In our numerical work, we set $T_0=1$.
As we demonstrate later, our model reproduces the quantum results
of Table I under the hypothesis that the time tags ${t}_{n ,1}$
are distributed uniformly over the interval
$[0,|\sin 2(\xi _{n } -\theta_1)|^d]$ with $d=2$.
Needless to say, we do not claim that our choice is the only one
that reproduces the results of quantum theory for the EPRB experiments.

\begin{figure*}[t] 
\begin{center}
\mbox{
\includegraphics[width=7.75cm]{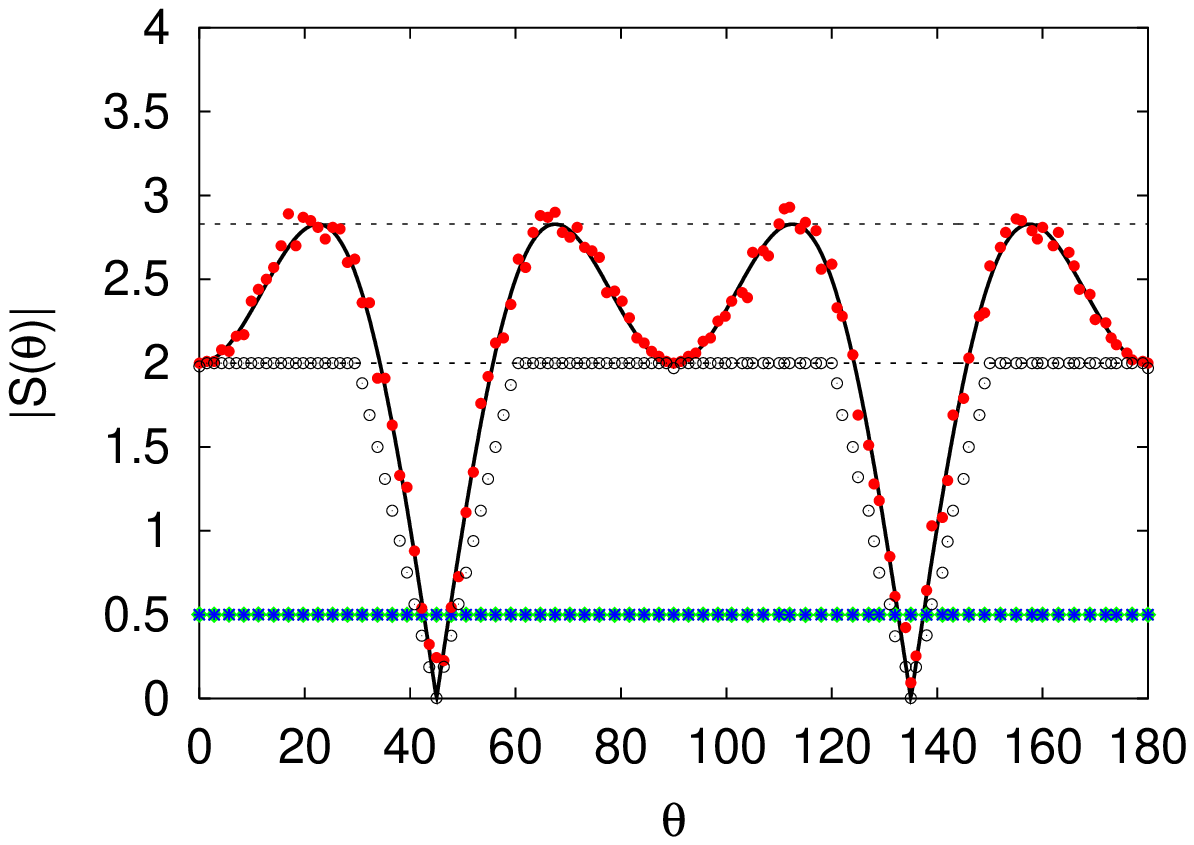}
\includegraphics[width=7.75cm]{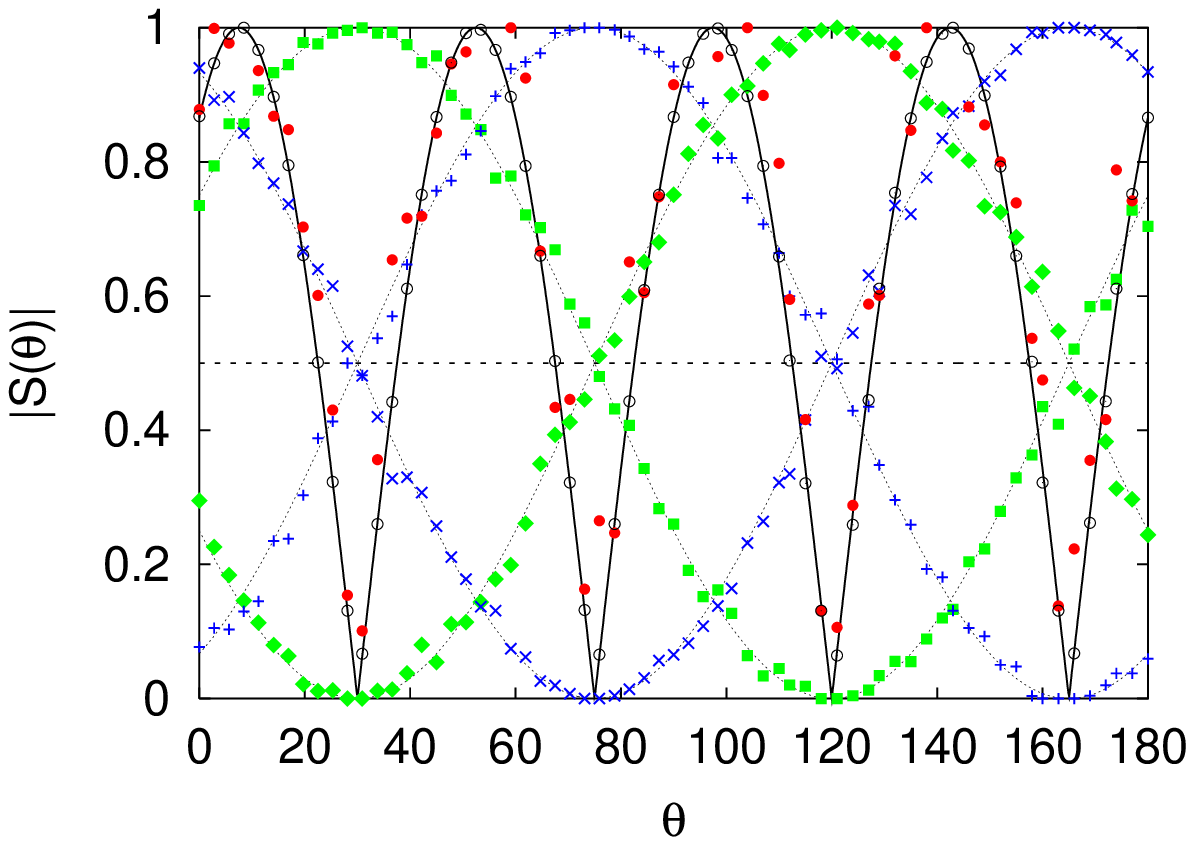}
}
\caption{(color online)
Left: Computer simulation of Experiment I in which the
source emits photons with opposite random polarization (EPRB experiment).
Right: Computer simulation of Experiment II in which the source emits photons with fixed polarization.
Solid circles (red): Simulation results using DLMs with the time-delay mechanism ($d=2)$ for the polarizers.
Open circles (black): Simulation results using DLMs but without using the time-tags (equivalent to
$d=0$ or $W\rightarrow\infty$) to compute the two-particle correlation.
Other markers: Average single-particle counts on the detectors (see Fig.~\ref{aspect}).
Squares (green): $P_{+}(\theta_1)=P_{+}(\theta^\prime_1)$;
Diamonds (green): $P_{-}(\theta_1)=P_{-}(\theta^\prime_1)$;
Plusses (blue): $P_{+}(\theta_2)=P_{+}(\theta^\prime_2)$;
Crosses (blue): $P_{-}(\theta_2)=P_{-}(\theta^\prime_2)$.
In Experiment I (left), these four symbols lie on top of each other.
In Experiment II (right), these markers show the typical Malus law behavior.
Solid line: Quantum theory for $S(\theta)$.
Dashed line at $S(\theta )=2\sqrt 2 $: Maximum of $S(\theta )$ if the system is described by quantum theory.
Dashed line at $S(\theta )=2$: Maximum of $S(\theta )$ if the system is described by the
class of models introduced by Bell~\cite{BELL93};
Dashed line at $S(\theta)=1/2$: Expected number of $+1$ and $-1$ events recorded by the detectors
if the input to the polarizers consist of photons with random polarization.
Dotted lines: Quantum theory for $P_{+}(\theta_1)=P_{+}(\theta^\prime_1)$, $P_{-}(\theta_1)=P_{-}(\theta^\prime_1)$,
$P_{+}(\theta_2)=P_{+}(\theta^\prime_2)$ and $P_{-}(\theta_2)=P_{-}(\theta^\prime_2)$.
}
\label{fig11}
\end{center}
\end{figure*}

\begin{figure*}[t] 
\begin{center}
\mbox{
\includegraphics[width=7.75cm]{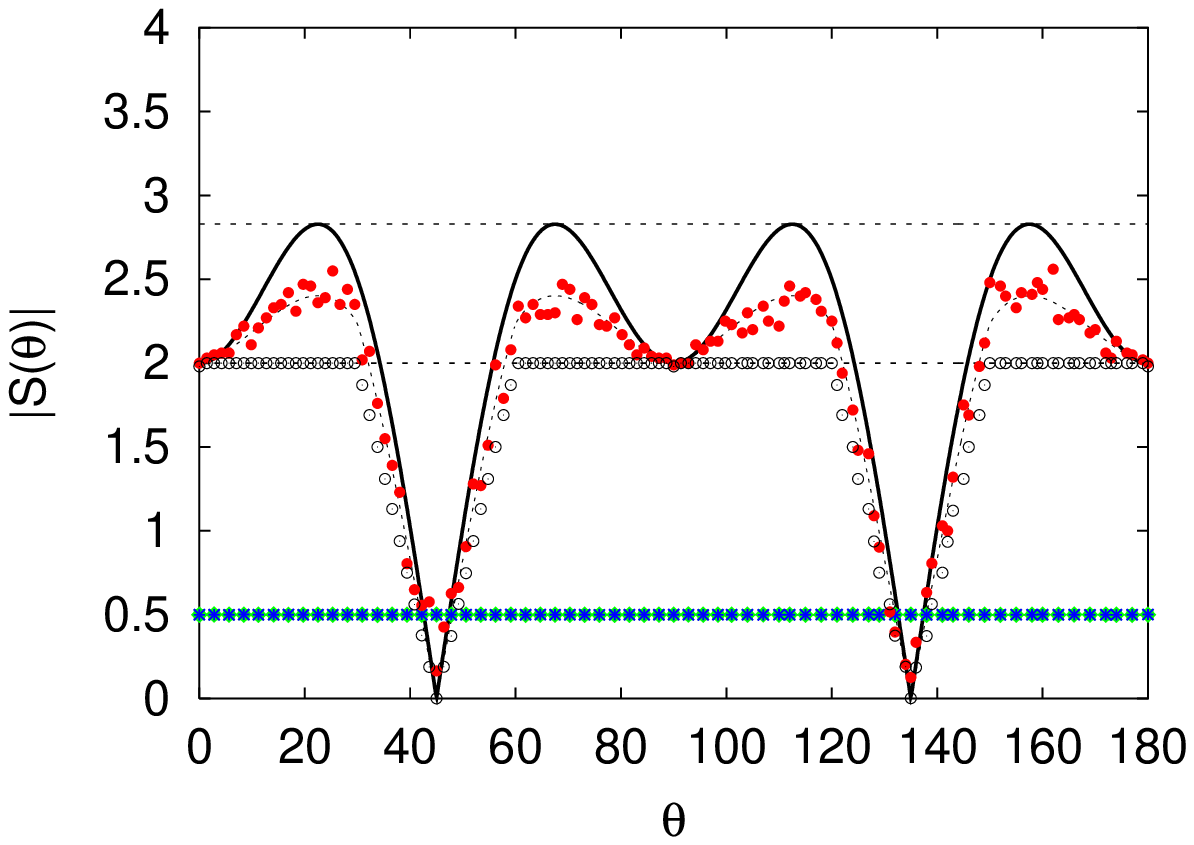}
\includegraphics[width=7.75cm]{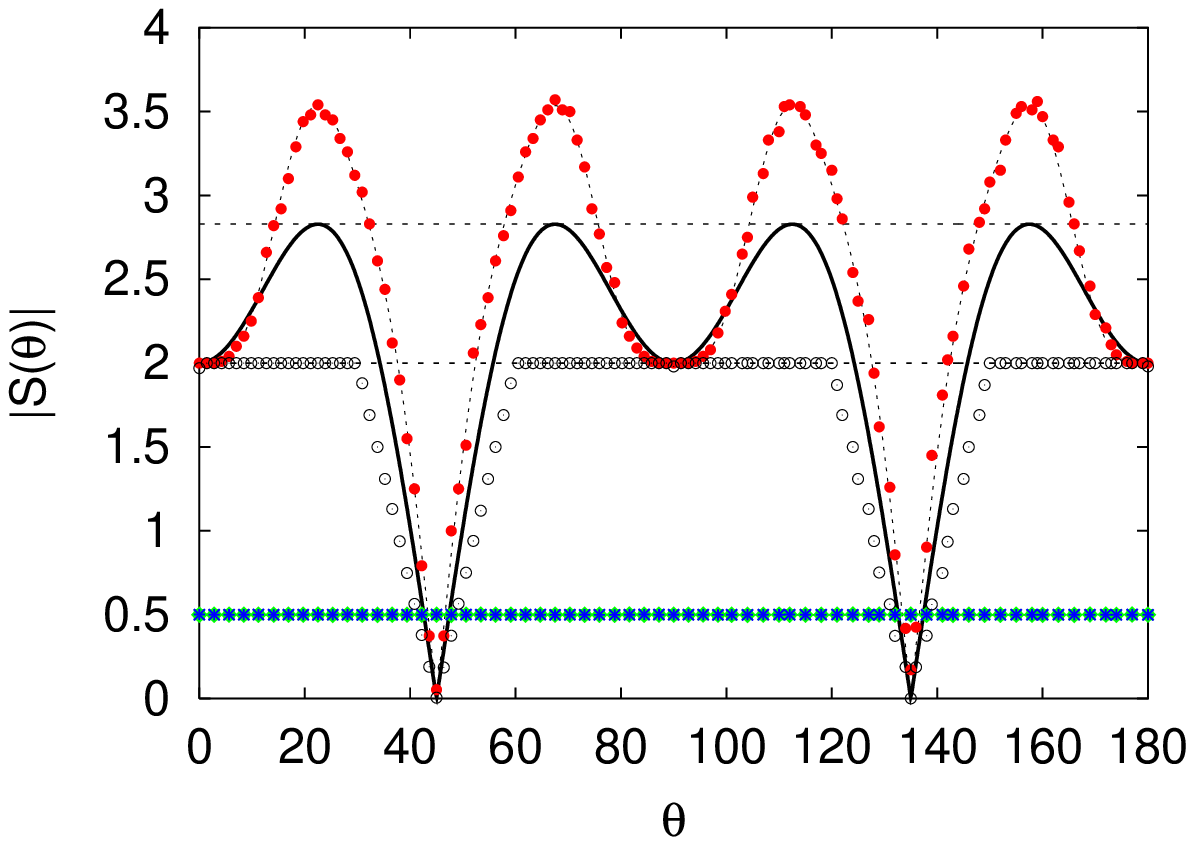}
}
\caption{(color online)
Left: Same as Fig.~\ref{fig11} (left) except that $d=1$.
Dotted line: $S(\theta)$ calculated from Eq.(\ref{Eabd1}).
Right: Same as Fig.~\ref{fig11} (left) except that $d=4$.
Dotted line: $S(\theta)$ calculated from Eq.(\ref{Eabd4}).
}
\label{fig12}
\end{center}
\end{figure*}

A last point left to address on the algorithm for the polarizer is
the deterministic character of the output
sequences generated by the DLMs. In fact, it is easy to change this
deterministic process into a random process without changing the
distribution of $+1$ and $-1$ events. A simple way to do this is to attach to
the particles, a random bit $z_n =\pm 1$, generated by the source. Then, we
modify the algorithm for the polarizer such that instead of selecting output
channel $\Delta $, it selects output channel $z_n\Delta $. This
procedure has no effect on the DLM dynamics and the single- and two-particle
counts but, it renders the outputs of the detectors unpredictable.
This technical finesse does not affect the final result for the expectation values and therefore
we disregard it in what follows.

\subsection{Data analysis}

For fixed $N$ and fixed angles $\theta_i$ and $\theta_i^{\prime}$ of the polarizers in the
observation stations, the algorithm described earlier generates the data sets $\Upsilon_i$,
just as experiment does. In order to count the coincidences, we choose a time-tag resolution $\tau$
and a time window $W$ such that $0<\tau<T_0$ and $\tau\le W$.
We set the single particle counts $P_x(\alpha )$, $P_y(\beta )$ and the coincidence counts
$C_{xy}(\alpha,\beta)$ with $\alpha=\theta_1,\theta_1^{\prime}$ and $\beta=\theta_2,\theta_2^{\prime}$,
to zero for all $x,y=\pm 1$.
Then, we make a loop over all events $x_{n,i}=\pm 1$
in the data sets and we read off $\gamma_{n,i}=\theta_{n,i},\theta_{n,i}^{\prime}$.
To count the coincidences, we first compute the discretized time tags $k_{n ,i} =\lceil t_{n ,i}/ \tau\rceil $
for all events in both data sets.
Here $\lceil{x}\rceil$ denotes the smallest integer that is larger or equal to $x$, that is
$\lceil{x}\rceil-1<x\le\lceil{x}\rceil$.
According to the procedure adopted in the experiment~\cite{WEIH98},
an entangled photon pair is observed if and only if
$\left| {k_{n,1} -k_{n,2} } \right|<k=\lceil{W/\tau}\rceil$.
Thus, if $\left| {k_{n,1} -k_{n,2} } \right|<k$,
we increment the count $C_{x_{n,1},x_{n,2}} (\alpha ,\beta )$ and we
increment the corresponding single particle counts $P_{x_{n,i}}(\gamma_{n,i})$.

We emphasize that the simulation procedure counts
all events that, according to the same criterion as the one employed in experiment,
correspond to the detection of single two-particle systems.


\section{Simulation results}\label{sec5}
\label{sec:results}

We use the computer model, described earlier to simulate the
experiment depicted in Fig.~\ref{aspect}.
Each polarizer in Fig.~\ref{aspect} is simulated by the same algorithm.
The procedure to direct the particles to the polarizers is the same
as in the laboratory experiment~\cite{ASPE82b}.
For each particle that enters station 1 (2),
a random number generator at station 1 (2) determines
which of the two polarizers will receive the particle.
In practice, we use two different random number generators
for station 1 and 2 (we have never seen any statistically
significant effect of using the same one for both stations).
The source always sends out two particles with orthogonal,
random polarization.
Unless we insert additional polarizers between
the source and stations 1 and 2, this setup simulates
Experiment I.
Inserting additional polarizers between
the source and stations 1 and 2 is the same as sending
the particles with fixed polarization to stations 1 and 2.
In this case, we simulate Experiment II.

The simulation proceeds in the same way as in the experiment, that is we first collect
the data sets $\Upsilon_1$ and $\Upsilon_2$ for various settings of the polarizers
(various $\gamma_{n,i}$), and then compute the single particle counts Eq. (\ref{eq11}),
the coincidences Eq.(\ref{Cxy}) and the
correlation Eq.(\ref{Exy}), from which we can calculate the function $S(\theta)$
(see Eq.(\ref{eq31})).

In Fig.~\ref{fig11} (left), we present our simulation data for Experiment I, that is
for the case that the source emits particles with an opposite, random
polarization, corresponding to the singlet state in the quantum mechanical
description.
The parameters in these simulations are $k=1$, $d=0,2$, $\tau=0.00025$, $l=0.999$, and $N=10^6$.
The results are not sensitive to the choice of these parameters.
For instance, the figures (not shown) with the results for $k=1$, $d=0,2$, $\tau=0.25$, $l=0.999$, and $N=10^6$
are barely distinguishable from those of Fig.2 and, as can be expected on general grounds,
increasing the number of events $N$ simply reduces the fluctuations.
The parameter $l$ controls the accuracy with which we can resolve differences in the angles:
The closer $l$ is to one, the higher the accuracy.
We have chosen $l$ such that, with the resolution used to plot the data, the effect of $l$
on the results cannot be noticed.
The main reason for showing the data for $k=1$, $d=0,2$, $\tau=0.00025$, $l=0.999$, and $N=10^6$
is that it allows us to demonstrate
that our simulation results are in excellent agreement with the analytical results of Appendix B.

It is clear that for $d=2$, the simulation model reproduces the results
of quantum theory for the single-particle expectation values $P_{\pm}(\alpha)$ and $P_{\pm}(\beta)$ (see Table I)
and $S(\theta)$ (see Eq.~(\ref{Stheta})).
Indeed, the frequency with which each detector fires is approximately one-half and
$S(\theta)$ agrees with the expression Eq.~(\ref{Stheta}) for the singlet state.
Also shown in Fig.~\ref{fig11} (left) are the results for $S(\theta)$
if we disable the time-delay mechanism.
Effectively, this is the same as letting the time window $W\rightarrow\infty$ or
setting $d=0$. Then, our simulation model generates data that
satisfies $|S(\theta)|\le2$, which is what we expect for the
class of models studied by Bell~\cite{BELL93}.

In Experiment II, the source emits particles with a fixed (not necessarily opposite)
polarization. In the right panel of Fig.~\ref{fig11}, we present results, obtained
by the same simulation algorithm as for Experiment I,
for the case $\theta_1=\theta_1^{\prime}=\alpha ={\alpha }'=\theta $ and
$\theta_2=\theta_2^{\prime}=\beta={\beta }'=\theta +\pi /4$.
The angle $\xi $ of the particles is $\pi /6$
(corresponding to $\eta_1=\pi /6$ and $\eta_2=\pi /6+\pi /2$ in the quantum mechanical description).
For this choice, we have $P_+ (\alpha)=\cos ^2(\theta -\pi/6 )$,
$P_+ (\beta)=\cos ^2(\theta -\pi /6 -\pi /4)$, $E(\alpha,\beta)=2^{-1}\sin
4(\pi /6-\theta)$ and $S(\theta )=\sin 4(\pi /6-\theta)$.
Except for the properties of the particles, the model
parameters for Experiment I and II are the same.

From Fig.~\ref{fig11}, it is clear that the event-by-event simulation
reproduces the single- and two-particle results of quantum theory for both
Experiment I and II, {\sl without any change to the algorithm that simulates the
polarizers}.

Having established that the data generated by our ``non-quantum'' system
agrees with quantum theory, it is of interest to explore if
these dynamical, adaptive systems can generate data that is not described by
quantum theory or by the simple, locally causal probabilistic models introduced by Bell~\cite{BELL93}.
We can readily give an affirmative answer to this question by
repeating the simulations for Experiment I (see Fig.~\ref{fig11} (left)) for different
values of the time-delay parameter $d$, all other parameters being the same as
those used to obtain the data presented in Fig.~\ref{fig11}.

For $d=0$, simulations with or without time-delay mechanism yields data
that, within the usual statistical errors, are the same (results not shown)
and satisfy $|S(\theta)|\le2$ .
Figure~\ref{fig12} shows the simulation data for $d=1$ and $d=4$. For $0<d<2$ our model yields
two-particle correlations that are stronger than those of the Bell-type
models but they are weaker than in the case of the singlet state in quantum theory.
Therefore, the maximum of $S(\theta )$ is less than $2\sqrt 2$ but larger than two.
For $d\ge3$, we find that the two-particle correlations are
significantly stronger than in the case of the singlet state in quantum theory.
From Fig.~\ref{fig12} it can be seen that for $d=1$ and $d=4$ there is
good agreement between the results obtained with our event-based simulation model and
the analytical result for $|S(\theta )|$ obtained from Eqs. (\ref{Eabd1}) and (\ref{Eabd4}), respectively (see Appendix B).
For $d=1$, the simulation results show larger fluctuations than for $d=4$, but in all cases they can be
reduced by increasing $N$ (results not shown).

\begin{figure}[t]
\begin{center}
\mbox{
\includegraphics[width=10cm]{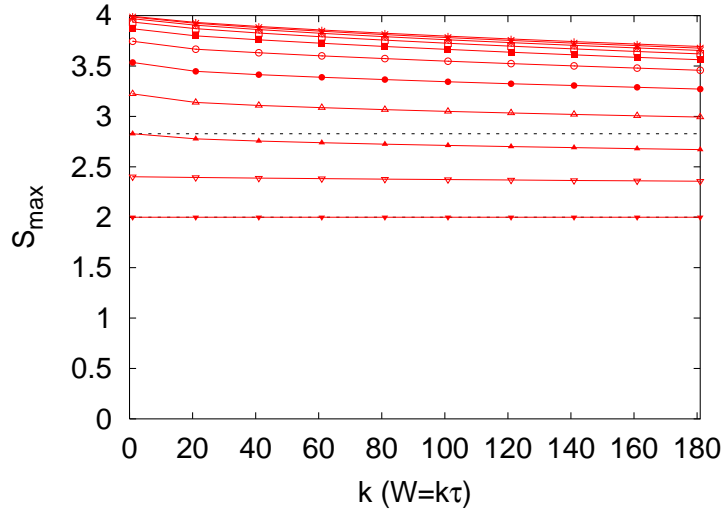}
}
\caption{(color online)
$S_{max}=\max_{\alpha,\alpha',\beta,\beta'} S(\alpha,\alpha',\beta,\beta')$
as a function of the time window $W$ relative to the time-tag resolution $\tau$.
Curves from bottom to top: Results for $d=0, 1, \ldots , 10$.
Dashed line: Value of $S_{max}=2\sqrt{2}$ if the system is described by quantum theory.
}
\label{maxStheta}
\end{center}
\end{figure}

The simulation results presented in Figs.~\ref{fig11} and \ref{fig12}
have been obtained for $W/\tau=1$ and small $\tau$ (recall that the unit of time
in our numerical work is set equal to one).
In general, in experiment the two-particle correlation depends on both $W$ and $\tau$.
Our simulation model makes definite predictions for this dependence.
This can be seen from Fig.~\ref{maxStheta} which shows
$S_{max}=\max_{\alpha,\alpha',\beta,\beta'} S(\alpha,\alpha',\beta,\beta')$
as a function of $W/\tau$ for
various values of $d$.
$S_{max}$ is calculated numerically using Eqs.~(\ref{EabNinfinity}) and (\ref{CK}) (see Appendix B).
The numerical results agree with the values of $S_{max}$ that have been obtained analytically
for $W=\tau\rightarrow 0$, $d=0,2$ and $W\rightarrow\infty$.
For $d<2$, $2\le S_{max}<2\sqrt{2}$ for any value of $W/\tau$. Hence, for $d<2$ our model cannot produce the
correlations of the singlet state.
For $d=2$, $2\le S_{max}\le 2\sqrt{2}$ and our model produces the correlations of the singlet state if $W/\tau\rightarrow 0$.
For $d>2$, $2\le S_{max}\le 4$, and for a range of $W/\tau$, $S_{max}>2\sqrt{2}$, implying that our model
exhibits correlations that cannot be described by the quantum theory of two spin-1/2 particles.

\section{Discussion}\label{sec6}
\label{sec:discussion}

We have pointed out that a violation of Bell's inequality is not an absolute
criterion for having quantum correlations.
The fact that our event-based model can produce correlations that violate the Bell inequality,
by itself, is not a surprise, because the expression for the coincidences Eq.~(\ref{Cxy})
cannot be written in a form that allows a derivation of the Bell inequality~\cite{LARS04}.
Given the two data sets, recorded at the two observation stations as described in Section~\ref{sec2},
we may find correlations between the data of both sets that may or may not violate the Bell inequality:
Mathematically such correlations cannot be excluded.~\cite{LARS04}
Our results do not contradict the folklore about Bell's theorem.
Bell's notion of locality is an attempt to incorporate Einstein's criteria of local causality
on the ontological level in probabilistic theories~\cite{JAYN89}.
In fact, in the derivation of Bell's inequality a strong mathematical (in terms of
probability theory) assumption was made, namely that any logical relation between both data sets is prohibited.
Physically, this assumption is usually erronously associated with the independence of operations at distant positions.
Thus the presence of any physical relation, which could be classical or quantum in origin, can lead
to a violation of the bounds in Bell's inequality.

The time-delay mechanism is an example of such a physical relation.
In our event-based model, the expression for the coincidences is the key ingredient
to reproduce the quantum mechanical results for the two-particle correlation of the EPRB experiment.
This expression, based on the time tags of the detection events,
is the same as the one employed in EPRB laboratory experiments with photons.
With this example, we may say that a simulation model that strictly satisfies
Einstein's criteria of local causality can reproduce the quantum theoretical results for EPRB experiments,
without using any concept from quantum theory.
That is, although our event generating and measurement processes are of classical origin,
they still lead to a violation of the original Bell inequality.
This fact must be considered seriously in any real experiment.
Instead of the time-delay, we may consider various other mechanisms which can cause a similar effect.
In fact, there may be many "unpaired signals".
This means that many photons are destroyed by some reason.
So far the destruction is considered to happen randomly.
But if it would depend on the relative angle between the photon polarization an the direction of the polarizer,
similar effects as those studied in the present paper could occur.
In any case, we have to be careful in eliminating those possibilities while studying the appearance of
quantum correlations. The dependence on the time window is a good check of this fact.

We have shown that in our model, in the case of Experiment I, the two-spin correlation depends on the value of the time window $W$.
By reducing $W$ from infinity to zero, this correlation changes from typical Bell-like to singlet-like,
without changing the procedure by which the particles are emitted by the source.
Thus, the character of the correlation not only depends on the whole experimental setup
but also on the way the data analysis is carried out.
Hence, from the two-spin correlation itself, one cannot make any definite statement about the character of the source.
Thus, the spin correlation is a property of the whole system
(which is what quantum theory describes), not a property of the source itself.
It is of interest to note that if we perform a simulation of Experiment II
the single-spin and two-spin correlations do not depend on the value of the time window $W$.
In this case, the observation stations always receive particles with the same polarization and
although the number of coincidences decreases with $W$ (and the statistical errors increase),
the functional form of the correlation does not depend on $W$.

We have also presented a rigorous proof that our simulation model reproduces the two-spin
correlation that is characteristic for the singlet state.
Furthermore, our model also allows us to explore phenomena
that cannot be described by quantum theory of two $S=1/2$ particles.


Finally, we also examined the effect that the pair identification criterion has on the two-particle correlations
for a set of experimental data that is publicly available. The results, presented in Appendix A, show a tendency
that is similar to the predictions of our simulation model, namely that the time window, used as a criterion to
identify photon pairs based on the time-tag data of single photon events, should be chosen as small as possible in order
to find results for the single-particle counts and two-particle correlations that agree
with the quantum theoretical expectation values for a system of two $S=1/2$ particles.

\section*{Appendix A}
\label{sec:appendixA}

We illustrate the importance of the choice of the time window
$W$ by analyzing a data set (the archives
Alice.zip and Bob.zip) of an EPRB experiment with photons
that is publicly available~\cite{WEIHdownload}.
Technically, the experiment of Ref.~\cite{WEIH98,WEIH00}
is different from the one sketched in Fig.~\ref{fig1},
but conceptually both experiments are the same.
The data in the archives Alice.zip and Bob.zip
are data for Experiment I in which $\theta_1=0$,
$\theta^\prime_1=\pi/4$,
$\theta_2=\pi/8$,
and
$\theta^\prime_2=3\pi/8$.

In the real experiment, the number of events detected at station 1 is unlikely
to be the same as the number of events detected at station 2.
In fact, the data sets of Ref.~\cite{WEIHdownload} show that
station 1 (Alice.zip) recorded $N_1=388455$ events while
station 2 (Bob.zip) recorded  $N_2=302271$ events.
Furthermore, in the real EPRB experiment, there may be an
unknown shift $\Delta$ (assumed to be constant during the experiment)
between the times $\{t_{n,1}|n=1,\cdots ,N_1\}$ gathered at station 1 and
the times $\{t_{m,2}|m=1,\cdots ,N_2\}$ recorded at station 2.
Therefore, there is some extra ambiguity in
matching the data of station 1 to the data of station 2.

A simple data processing procedure that resolves this
ambiguity consists of two steps~\cite{WEIH00}.
First, we make a histogram of the time differences
$t_{n,1}-t_{m,2}$ with a small but reasonable resolution
(we used $0.5$ ns).
Then, we fix the value of the time-shift $\Delta$
by searching for the time difference for which
the histogram reaches its maximum, that is we maximize
the number of coincidences by a suitable choice of $\Delta$.
For the case at hand, we find $\Delta=4$ ns.
Finally, we compute the coincidences,
the two-particle average, and $S_{max}$ using
the same expressions as the ones used to analyze the computer simulation data.
The average times between two detection events is $2.5$ ms and $3.3$ ms
for Alice and Bob, respectively.
The number of coincidences (with double counts removed) is
13975 and 2899 for ($\Delta=4$ ns, $W=2$ ns) and
($\Delta=0$ , $W=3$ ns) respectively.

\begin{figure}[t]
\begin{center}
\includegraphics[width=10cm]{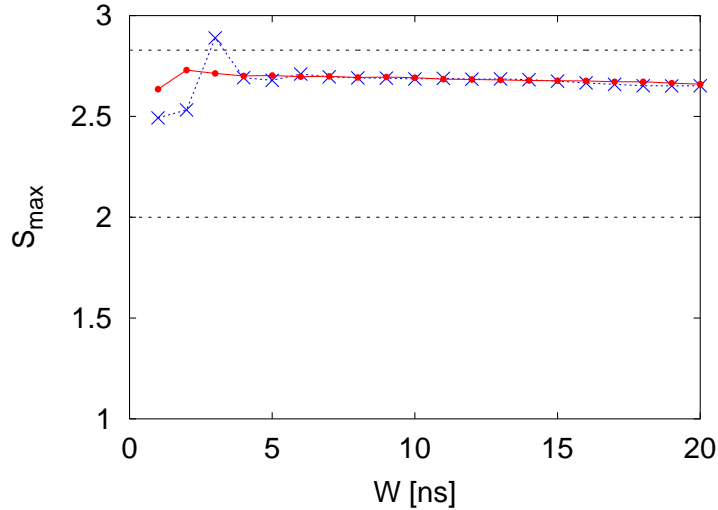}
\caption{(color online) $S_{max}$ as a function of the time window $W$,
computed from the data sets contained in the archives
Alice.zip and Bob.zip that can be downloaded from Ref.~\cite{WEIHdownload}.
Bullets (red): Data obtained by
using the relative time shift $\Delta=4$ ns that maximizes the
number of coincidences.
The maximum value of $S_{max}\approx2.73$ is found at $W=2$ ns.
Crosses (blue): Raw data ($\Delta=0$).
The maximum value of $S_{max}\approx2.89$ is found at $W=3$ ns.
Dashed line at $|S(\theta )|=2\sqrt 2 $: $S_{max}$ if the system is described by quantum theory (see Section~\ref{sec:quantum}).
Dashed line at $|S(\theta )|=2$: $S_{max}$ if the system is described by the
class of models introduced by Bell~\cite{BELL93}.
}
\label{exp2}
\end{center}
\end{figure}

In Fig.~\ref{exp2} we present the results
for $S_{max}$ as a function of the time window $W$.
First, it is clear that $S_{max}$ decreases as $W$ increases.
Second, the procedure of maximizing the coincidence count
by varying $\Delta$ reduces the maximum value of $S_{max}$ from
a value 2.89 ($\Delta =0$) that considerably exceeds the maximum
for the quantum system ($2\sqrt{2}$, see Section~\ref{sec:quantum}) to a value 2.73
(the value cited in Ref.~\cite{WEIH98})
that violates the Bell inequality and is less than the maximum for the quantum system.
The optimized experimental results (bullets in Fig.~\ref{exp2}) and
the results of our simulation model (see Fig.~\ref{maxStheta}, third line from the bottom ) are qualitatively very similar.

The fact that the ``uncorrected'' data ($\Delta=0$) violate the rigorous bound for the quantum system
should not been taken as evidence that quantum theory is ``wrong'':
It merely indicates that the way in which the data of the two stations
has been grouped in two-particle events is not optimal.

\begin{figure}[t]
\begin{center}
\includegraphics[width=10cm]{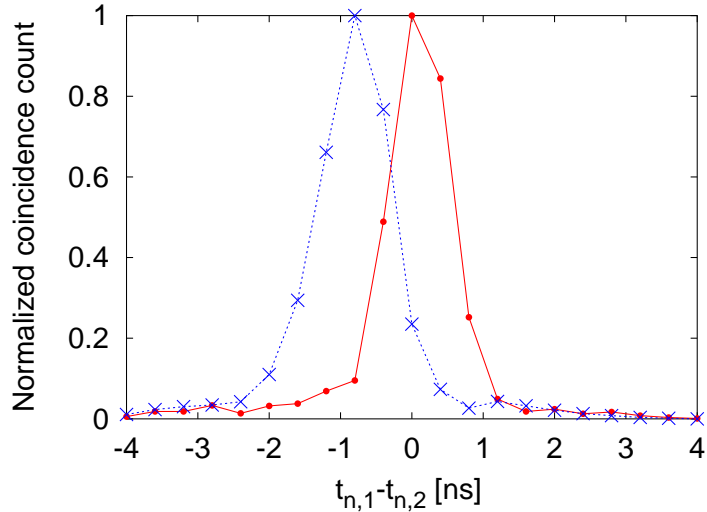}
\caption{(color online) Normalized coincidence counts as a function of time tag difference $t_{n,1}-t_{n,2}$,
computed from the data sets contained in the archives
Alice.zip and Bob.zip~\cite{WEIHdownload}, using the relative time shift $\Delta=4$ ns that maximizes the
number of coincidences.
Bullets (red): $\theta_1=0$ and $\theta_2=\pi/8$;
Crosses (blue): $\theta_1=0$ and $\theta_2=3\pi/8$.
}
\label{fort.7}
\end{center}
\end{figure}

Finally, we use the experimental data to show that the time delays depend on the
orientation of the polarizer. To this end, we
select all coincidences between $D_{+,1}$ and $D_{+,2}$ (see Fig.~\ref{fig1}) and
make a histogram of the coincidence counts as a function of the time-tag difference,
for fixed orientation $\theta_1=0$ and the two orientations $\theta_2=\pi/8,3\pi/8$ (other
combinations give similar results).
The results of this analysis are shown in Fig.~\ref{fort.7}.
The maximum of the distribution shifts by approximately 1 ns as the polarizer at station 2 is
rotated by $\pi/4$, a demonstration that the time-tag data is sensitive to the orientation
of the polarizer at station 2. A similar distribution of time-delays (of about the same width) was also observed in a much older
experimental realization of the EPRB experiment~\cite{KOCH67}.
The birefringent properties of the optical elements (polarizers and electro-optic modulators)
might be responsible for this time delay. A more detailed quantitative and exploratory analysis
of this time delay requires dedicated retardation measurements for these specific optical elements
in single-photon set-ups.

\begin{figure}[t]
\begin{center}
\mbox{
\includegraphics[width=10cm]{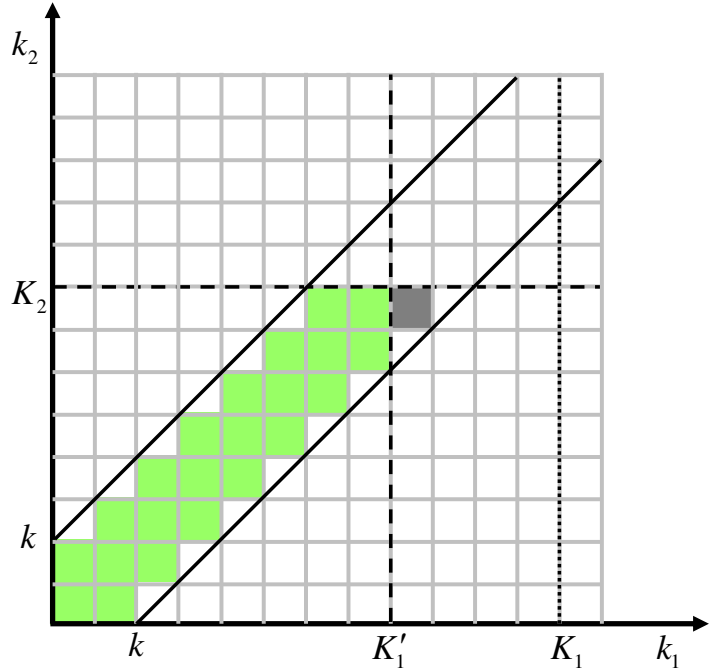}
}
\caption{(color online)
Graphical representation of the process of counting pairs. The time
interval is divided in bins of size \textit{$\tau $}, represented by the elementary squares.
The two parallel, 45$^{o}$ lines indicate the time window $W$, which was chosen
to be 2\textit{$\tau $} in this example. In the limit $N\rightarrow \infty $, the total number of
pairs for fixed $\alpha$, $\beta$, and $\xi$
is given by the number of whole squares that fall within the time
window and satisfy $1\le k_i <K_i $ for $i=1,2$. For $K_1 >K_2 $, all filled
squares contribute while for ${K}'_1 =K_2 $, the dark gray square does not
contribute. For $K_1<K_2$ we interchange labels 1 and 2.
}
\label{math}
\end{center}
\end{figure}

\section*{Appendix B}
\label{sec:appendix}
In the case of Experiment I and for some choices of the model parameters,
we can compute the correlation Eq.~(\ref{Exy}) analytically~\cite{RAED06c}.
In the limit $N\rightarrow\infty$, Eq.~(\ref{Exy}) can be written as
\begin{equation}
\label{EabNinfinity}
E(\alpha,\beta)=-\frac{\int_0^{2\pi }x_1(\xi,\alpha)x_2(\xi,\beta)P(T_1,T_2,W)d\xi}{\int_0^{2\pi }P(T_1,T_2,W) d\xi},
\end{equation}
%
where $P(T_1,T_2,W)$ is the density
of coincidences for fixed $(\alpha ,\beta )$ and polarization angle $\xi$ (within a small interval $d\xi$),
$T_1=|\sin2(\xi-\alpha)|^d$,
$T_2=|\sin2(\xi-\beta)|^d$,
$x_1(\xi,\alpha)=\mathop{\hbox{sign}}(\cos2(\xi-\alpha))$,
and $x_2(\xi,\beta)=\mathop{\hbox{sign}}(\cos2(\xi-\beta))$.

Mathematically, expression Eq.~(\ref{EabNinfinity}) can give rise to almost any correlation.
The density of coincidences 
appears as a result of integrating over the
distribution of time tags and it is this integration that leads to the appearance of the correlations.
Of course, the specific form of the correlations depends on the choice of the functional dependence of $T_1$ and
$T_2$ on the angles.

The expression for $P(T_1,T_2,W)$ can be derived as follows.
For a fixed time-tag resolution $0<\tau <1$, the discretized time-tag
for the $n$th detection event is given by $k_{n,i}
=\left\lceil {t_{n,i} \tau ^{-1}} \right\rceil $ where $\left\lceil x
\right\rceil $ denotes the smallest integer that is larger or equal to $x$.
The discretized time-tag $k_{n,i}$ takes integer values between 1 and $K_i
\equiv\lceil \tau ^{-1}T_i \rceil $, where $K_i$ is the
maximum, discretized time delay for a particle with polarization $\xi$
and passing through the polarizer with
orientation $\gamma_i$, where $\gamma_1=\alpha$ and $\gamma_2=\beta$.
If $\vert k_{n,1}-k_{n,2} \vert <k=\left\lceil {\tau ^{-1}W} \right\rceil $,
the two photons are defined to form a pair.
For fixed $\alpha$, $\beta$, and $\xi$, we can count the total number of pairs,
or coincidences $C$, by considering the graphical representation
shown in Fig.~\ref{math}.
After a careful examination of all possibilities, we find that
\begin{align}
 C\equiv C(K_1 ,K_2 ,k)&= (2k_0 -1)k_{12} -k_0 (k_0 -1)/2
\nonumber \\
 &- \max (0,(K_{12}-1)\max (0,K_{12} )/2)
\nonumber \\
& + \max (0,k-k_0 )k_0
\nonumber \\
& -\max (0,kk_{12} -K_1 K_2 )
,
\label{CK}
\end{align}
where $k_0 =\min (K_1 ,K_2 ,k)$, $k_{12} =\min (K_1 ,K_2 )$, and $K_{12}
=k_{12} -\max (0,\max (K_1 ,K_2 )-k)$.

It is clear that the result for the coincidences depends on the
time-tag resolution $\tau$, the time window $W$ and the number of events $N$,
just as in real experiments~\cite{FREE72,ASPE82b,ASPE82a,TAPS94,TITT98,WEIH98,ROWE01,FATA04}.
%
Formula Eq.~(\ref{CK}) greatly simplifies
if we consider the case $k=1$ ($W=\tau$), yielding
$C(K_1 ,K_2 ,1)=\min (K_1 ,K_2 )$
as is evident by looking at Fig.~\ref{math}.
For fixed $\alpha$, $\beta$, and $\xi$,
and $W=\tau$, the density $P(T_1,T_2,\tau)=C(K_1 ,K_2
,1)/K_1 K_2 $ that we register two particles with a time-tag difference less than $\mathit{\tau }$ is bounded by
\begin{equation}
\tau \frac{\min (T_1+\tau ,T_2+\tau )}{( T_1+\tau)( T_2+\tau )}<P(T_1 ,T_2 ,\tau)\le\tau\frac{\min (T_1,T_2)}{T_1T_2}.
\label{CK1bound}
\end{equation}
For $W=\tau\rightarrow 0$ and $d=2$, Eq.~(\ref{EabNinfinity}) reads
\begin{align}
\label{EabW0}
E(\alpha,\beta)&=-\frac{
\int_0^{2\pi} x_1(\xi,\alpha)x_2(\xi,\beta)\frac{\min(\sin^2(\xi-\alpha),\sin^2(\xi -\beta))}{\sin^2(\xi-\alpha)\sin^2(\xi -\beta)} d\xi
}{
\int_0^{2\pi} \frac{\min(\sin^2(\xi-\alpha),\sin^2(\xi -\beta))}{\sin^2(\xi-\alpha)\sin^2(\xi -\beta)} d\xi
}
\nonumber \\
&= -\cos2(\alpha-\beta),
\end{align}
in exact agreement with the quantum mechanical result (see Table I).

For $d=0$, $d=1$, and $d=4$ we find
\begin{equation}
\label{Eabd0}
E(\alpha,\beta )= -1+\frac{2\vert \alpha-\beta \vert \hbox{mod} \pi }{\pi },
\end{equation}
\begin{equation}
\label{Eabd1}
E(\alpha,\beta )=-\frac{\ln \frac{1+|\cos (\alpha -\beta )|}{1-|\cos (\alpha -\beta )|}
\frac{1-|\sin (\alpha -\beta )|}{1+|\sin (\alpha -\beta )|}}
{\ln \frac{1+|\cos (\alpha -\beta )|}{1-|\cos (\alpha -\beta)|}\frac{1+|\sin (\alpha -\beta )|}{1-|\sin (\alpha -\beta)|}},
\end{equation}
and
\begin{equation}
\label{Eabd4}
E(\alpha,\beta )=-\frac{(3-\cos^22(\alpha -\beta ))\cos 2(\alpha -\beta )}{2},
\end{equation}
respectively.
The corresponding results for $S(\theta )$ are shown in Fig.~\ref{fig12}.

If $W\rightarrow \infty$, $\Theta (W-\vert t_{n,1}-t_{n,2} \vert )=1$, and
$P(T_1,T_2,W)=1$ such that Eq.~(\ref{EabNinfinity}) reduces to~\cite{BELL93}
\begin{align}
\label{EabWinfinity}
E(\alpha,\beta)&=\frac{1}{2\pi} \int_0^{2\pi}
\mathop{\hbox{sign}}(\cos2(\xi-\alpha))\mathop{\hbox{sign}}(\cos2(\xi-\beta))d\xi
\nonumber \\
&= -1+\frac{2\vert \alpha-\beta \vert \hbox{mod} \pi }{\pi }
.
\end{align}
Obviously, Eq.~(\ref{EabWinfinity}) does not agree with the quantum theoretical expression
$E(\alpha,\beta)=-\cos2(\alpha-\beta)$.

%

\end{document}